\documentclass[twocolumn]{aastex62}

\usepackage{graphics,epsf}
\usepackage{amsmath}                
\usepackage{amsfonts}               
\usepackage{amssymb}                
\usepackage{epsfig}                 
\usepackage{float}
\usepackage{color}
\usepackage{tabularx}
\usepackage{comment}
\usepackage{makecell}

\def \eV{~\rm{eV}}

\def \s{~\rm{s}}
\def \km{~\rm{km}}

\def \yr{~\rm{yr}}
\def \Myr{~\rm{Myr}}
\def \Gyr{~\rm{Gyr}}

\usepackage{xcolor}
\definecolor{redak}{rgb}{0.9,0.15,0.05}

\begin{document}
\title{Common envelope jets supernova r-process yields can reproduce $\rm [Eu/Fe]$ abundance evolution in the Galaxy}

\author{Aldana Grichener}
\affiliation{Department of Physics, Technion, Haifa, 3200003, Israel; aldanag@campus.technion.ac.il; soker@physics.technion.ac.il}

\author{Chiaki Kobayashi}
\affiliation{Centre for Astrophysics Research, Department of Physics, Astronomy and Mathematics, University of Hertfordshire, Hatfield, AL10 9AB, UK; c.kobayashi@herts.ac.uk}

\author[0000-0003-0375-8987]{Noam Soker}
\affiliation{Department of Physics, Technion, Haifa, 3200003, Israel; aldanag@campus.technion.ac.il; soker@physics.technion.ac.il}

\begin{abstract}
We use a numerical Galactic chemical evolution model and find that the common envelope jets supernova (CEJSN) r-process scenario can account for both the very early average ratio of europium to iron and its evolution at later times in the Milky-Way (MW) Galaxy. In the CEJSN scenario a neutron star (NS) spirals-in inside a red supergiant (RSG) star all the way to the core and destroys it. According to this scenario r-process isotopes are nucleosynthesized inside neutron-rich jets that the accretion disk around the NS launches inside the core. The merger of a NS with an RSG core already takes place in the very young Galaxy. We conclude that CEJSNe can be a major contributor to r-process nucleosynthesis. 
\end{abstract}


\section{INTRODUCTION}
\label{sec:intro}
  
The formation sites of heavy elements is an unsolved problem in nuclear astrophysics. Various astrophysical objects and transient events have been studied as possible sources of heavy r-process elements over the past decades. Among the possible candidates are magnetorotational supernovae (e.g., \citealt{Winteleretal2012, nis15, HaleviMosta2018, Reichertetal2021, Yongetal2021}), binary neutron star (NS)  mergers (e.g., \citealt{Gorielyetal2011, wan14, Metzger2017, vandevoortetal2021, 
Dvorkinetal2021}) and collapsars (e.g.,  \citealt{Siegeletal2019, Braueretal2021, Siegeletal2021}). These astronomical sites have been tested with Galactic chemical evolution models that lead to different conclusions. Recent studies suggest that multiple r-process sites are required to explain the r-process abundances in the Galaxy (e.g., \citealt{Wehmeyeretal2015}; \citealt{hay19, Yamazakietal2021, Farouqietal2021, Tsujimoto2021, Naiduetal2021}). \cite{Moleroetal2021} argue that the same applies for ultra faint dwarf galaxies.However,the potential need for multiple dominant r-process sources is still under debate, and several studies argue that binary NS mergers dominate the r-process nucleosynthesis (e.g., \citealt{Jietal2016, Beniaminietal2016a, Beniaminietal2016b, Banerjeeetal2020}).  

Another r-process site that might contribute to the formation of heavy elements is the common envelope jets supernova (CEJSN) r-process scenario (\citealt{Papishetal2015, GrichenerSoker2019a, GrichenerSoker2019b}), in which r-process nucleosynthesis occurs in NS-red super giant (RSG) common envelope systems where a cold NS is engulfed by a RSG and orbits inside it. The NS accretes mass from the envelope of the RSG through an accretion disk and eventually reaches the CO core, where it accretes mass at a higher rate. The gravity of the NS destroys the core forming a thicker accretion disk around the NS, and the massive thick disk launches neutron-rich jets in which r-process nucleosynthesis can occur. 

The physics of CEJSNe and their possible outcomes are a topic of ongoing research (e.g., \citealt{MorenoMendezetal2017, Gilkisetal2019, LopezCamaraetal2019, LopezCamaraetal2020, Gricheneretal2021}). Due to their wide diversity CEJSNe events can account for various phenomena in astrophysics, such as peculiar supernovae whose light curves might be shaped by the operation of jets at different episodes during the explosion (e.g., \citealt{SokerGilkis2018, Sokeretal2019}). \cite{GrichenerSoker2021} argue that the ultra-relativistic jets that a black hole (BH) launches as it orbits in the envelope of a RSG can emit the high energy neutrinos ($\approx 10^{15} \eV$) that the IceCube neutrino observatory have been detecting since 2013 (e.g., \citealt{Aartsenetal2013}).

The evolution towards CEJSNe with a NS companion is almost identical to the evolution of binary systems that result in binary NS mergers. In both cases the scenario begins with two massive stars in a binary system. The more massive star (of $M \gtrsim 15 M_\odot$) explodes first as a core collapse supernova (CCSN) leaving a NS behind, while its companion (of $M \gtrsim 10 M_\odot$) evolves slower, and eventually reaches the RSG phase. The RSG star expands and swallows the NS due to tidal forces, initiating a common envelope evolution (CEE) phase in which the NS spirals in-inside the envelope of the giant star. If the entire envelope is ejected before the NS reaches the CO core, then the NS remains outside the core, and a NS-core binary system is formed. The core eventually explodes in a second CCSN event forming another NS. Both NSs might merge by emitting gravitational waves as their distance decreases (e.g., \citealt{VignaGomezetal2018}, \citealt{Fragosetal2019}). However, if the NS enters the core of the RSG before the core explodes the system will result in a CEJSN event. In \cite{GrichenerSoker2019a} we estimate that about one in ten events of a NS that enters the envelope of a giant star ends as a CEJSN.

In this \textit{letter} we study the contribution of CEJSNe to europium time evolution in the Galaxy. In section \ref{sec:Numerics} we describe our chemical evolution model along the parameter space of the CEJSN r-process scenario. In section \ref{sec:EuropiumTimeEvolution} we present our results and compare them to observations. We summarize in section \ref{sec:Summary}.

\section{The numerical scheme}
\label{sec:Numerics}

The Galactic chemical evolution code we use in this paper is the same as in \cite{Kobayashietal2020} but includes CEJSNe as the r-process site instead of magnetorotational supernovae and neutron star mergers. The slow neutron capture process and electron capture supernovae are included, but these do not contribute to the production of europium (see Fig. 32 of \citealt{Kobayashietal2020}). The code provides the production of elements following gas inflow and star formation in the solar neighborhood, assuming there is no outflow of gas out from the Galaxy. The timescales of the inflow and star formation are determined from independent observational constraints, e.g., the metallicity distribution function (see Fig. 2 of \citealt{Kobayashietal2020}), and the results do not depend so much on these parameters (see Fig. A1 of \citealt{Kobayashietal2020}). The resultant star formation history shows a peak at a few Gyrs ago. The natal kick of the NS is not included. The ISM is assumed to instantaneously mix, corresponding to an efficient turbulent mixing.

As in \cite{Kobayashietal2020}, Kroupa IMF is adopted for $0.01M_\odot$ to $50M_\odot$, and a significant amount of Fe is produced by CCSNe (hypernovae) in addition to Type Ia supernovae (SNe Ia). CCSNe dominate at early times, and the contribution of both CCSNe and SNe Ia to iron production becomes equal about 5 Gyrs after the formation of the MW. Afterwards, most of the iron is produced by SNe Ia. The nucleosynthesis yields and progenitor model of SNe Ia are taken from \citet{kob20ia}, which is based on Chandrasekhar explosions of C+O white dwarfs in the single degenerate scenario. The SNe Ia rate is calculated depending on metallicity. As a result the observed $\rm [O/Fe]-[Fe/H]$ relation is very well reproduced including a sharp decreasing trend from $\rm [Fe/H] \gtrsim -1$. We note though that other SN Ia scenarios exist (e.g., table 1 in \citealt{Soker2019}). 

The time-metallicity evolution of r-process elements depends strongly on the typical delay time from star formation to r-process nucleosynthesis by the events. For CEJSNe, this delay is mostly due to the time it takes to form the RSG-NS binary system, as the timescale of the CEE is much shorter. This delay time resembles the typical time to form a binary NS system since they both depend mainly on the stellar evolution time of massive stars (few to tens of $\Myr$; e.g., \citealt{Crowtheretal2012}). However, since the merger of the NSs is driven by radiation of gravitational waves, the additional delay until the merger is usually much longer (tens of $\Myr$ to millions of $\Gyr$; e.g., \citealt{BeniaminiPiran2019}). According to \cite{GrichenerSoker2019b} the typical delay time from star formation to r-process enrichment by CEJSNe is $t_{\rm d} \simeq 10-30\Myr$.

To find the europium time evolution in the Milky-Way (MW) due to CEJSNe we estimate the europium mass from a single event. The total mass of r-process yield per one CEJSN r-process event is $\rm 0.01-0.03 M_{\rm \odot}$ \citep{GrichenerSoker2019a}. We assume the solar abundance distribution of r-process elements. According to table 2 in \cite{Coteetal2018} the ratio of the mass fraction of europium to the total mass fraction of r-process elements in the solar r-process residuals is $\simeq 10^{-3}$. Therefore, for the CEJSN r-process yield estimated in \cite{GrichenerSoker2019a} the europium mass that one typical CEJSN event forms is $M_{\rm Eu} \simeq 1 \times \rm 10^{-5} - 3 \times10^{-5} M_{\rm \odot}$. From the population synthesis study performed by \cite{Schroderetal2020} we estimate the rate of CEJSNe with a NS companion to be about $300 \rm Gpc^{-3}\yr^{-1}$.

\section{The ${\rm [Eu/Fe]}$-${\rm [Fe/H]}$ evolution plane}
\label{sec:EuropiumTimeEvolution}

\subsection{Observations and possible sites} 
\label{subsec:ObservationsSites}

\begin{figure}
\begin{center}
\vspace*{-5.9cm}
\hspace*{-2.1cm}
\includegraphics[width=0.7\textwidth]{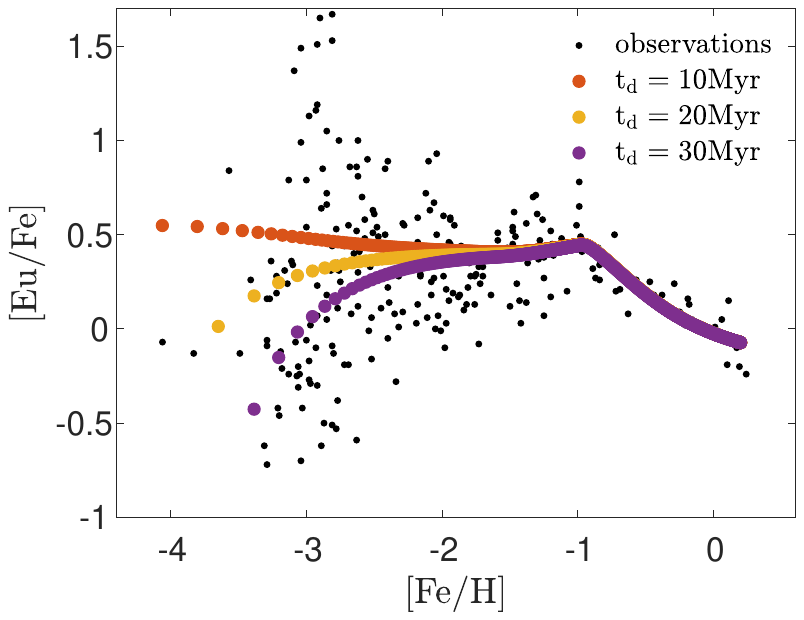}
\vspace*{-5.9cm}
\caption{The ${\rm [Eu/Fe]}$ over ${\rm [Fe/H]}$ distribution in the Milky-Way due to the contribution of common envelope jets supernovae that produce europium mass of $M_{\rm Eu}=2\times10^{-5} M_{\rm \odot}$ and have delay times of $t_{\rm d}=10\Myr$ (orange large dots), $t_{\rm d}=20\Myr$ (yellow large dots) and $t_{\rm d}=30\Myr$ (purple large dots). The black dots represent observational data taken from \cite{Cayreletal2004}, \cite{Hondaetal2004}, \cite{Hansenetal2012}, \cite{Hansenetal2014}, \cite{Roedereretal2014} and \cite{Zhaoetal2016}.     
}
\label{fig:DelayTimes}
\end{center}
\end{figure}

\begin{figure}
\begin{center}
\vspace*{-5.9cm}
\hspace*{-2.1cm}
\includegraphics[width=0.7\textwidth]{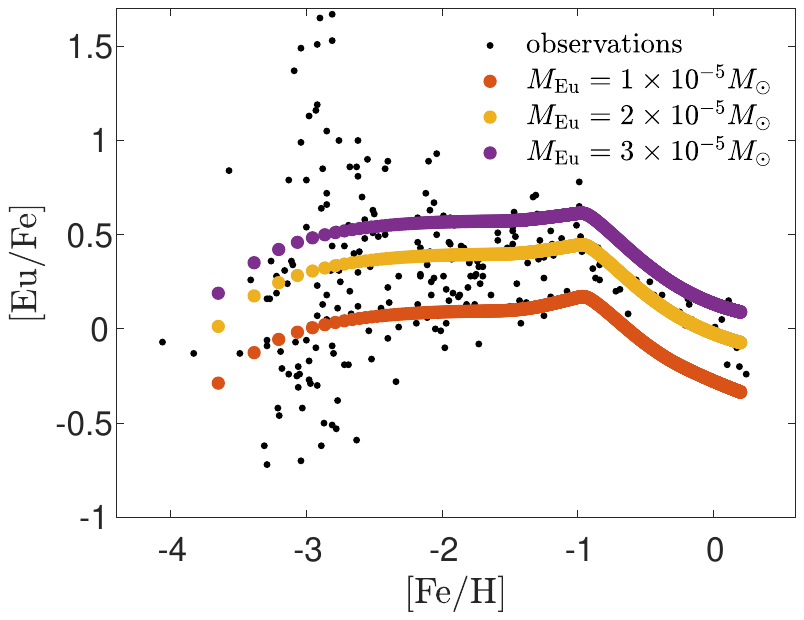}
\vspace*{-5.9cm}
\caption{Same as Fig. \ref{fig:DelayTimes} for common envelope jets supernovae with a delay time of $t_{\rm d}=20\Myr$ that produce europium mass of $M_{\rm Eu}=1 \times 10^{-5} M_{\rm \odot}$ (orange large dots), $M_{\rm Eu}=2\times10^{-5} M_{\rm \odot}$ (yellow large dots) and $M_{\rm Eu}=3\times10^{-5} M_{\rm \odot}$ (purple large dots). 
}
\label{fig:EuMass}
\end{center}
\end{figure}

Observations show that the ratio of europium to iron in the MW changes with metalicity. In Figs. \ref{fig:DelayTimes} and \ref{fig:EuMass} we present the observed values of $\rm [Eu/Fe]$ versus $\rm [Fe/H]$ for stars in the MW (black dots). We use the standard notation $\rm [X/Y]=log_{10} (N_{\rm X}/N_{\rm Y})-log_{10} (N_{\rm X}/N_{\rm Y})_{\rm \odot}$, where $N_{\rm X}$ and $N_{\rm Y}$ are the column density of element X and element Y, respectively. The observational data was selected carefully based on spectral resolution and abundance analysis methods (see \citealt{Kobayashietal2020} for more details). For older stars with lower metalicities of ${\rm [Fe/H]} \la -1$ there is a large scatter in the values of ${\rm [Eu/Fe]}$ around about the same average value of $\rm [Eu/Fe] \simeq 0.4$, while for stars with ${\rm [Fe/H] \ga -1}$ the ratio $\rm [Eu/Fe]$ decreases with increasing $\rm [Fe/H]$. This decrease is commonly known as the evolution `knee'. The transition between the two behaviours occurs for ${\rm [Fe/H]} \simeq -1$, corresponding to a Galaxy age of $\simeq 2 \Gyr$.


The large europium scatter observed at low metalicites suggests that r-process nucleosynthesis sites are rare (e.g., \citealt{Francoisetal2007}; \citealt{ BeniaminiHotokezaka2020}), and that the $\rm [Eu/Fe]$ chemical evolution pattern might have been produced by dilution of the gas that contained the heavy elements before the beginning of extensive iron production (e.g., \citealt{Shenetal2015}). This implies that at least one of the sites that contribute to r-process production should be spatially uncorrelated with iron sources. We note, however, that total un-correlation with iron results in too large scatter of $\rm [Eu/Fe]$ (e.g., \citealt{hay19}). 

The high natal kick velocities of the binary NS systems can bring them far from their birthplace before they merge, away from the location of their progenitor iron-producing CCSNe. However, \cite{Safarzadehetal2019b} found that binary NS mergers cannot account for r-process enrichment in the MW at early times given their long delay times (for another view see \citealt{BeniaminiPiran2019}; \citealt{Tarumietal2021}). Collapsars, on the other hand, do not have natal kicks and therefore the r-process elements they produce are likely to mix with the iron from the CCSN explosion, resulting in overproduction of r-process elements relative to iron in metal poor stars in the MW (e.g., \citealt{MaciasRamirez-Ruiz2019, FraserSchonrich2021}).

Binary NS mergers have a problem accounting for the ${\rm [Eu/Fe]}$ trend in the MW for higher metalicities as well. The commonly accepted delay time distribution expected from binary NS mergers (of $\propto t^{-1}$) cannot explain the evolution `knee', leading to the need of either a modification of the delay time distribution (e.g., \citealt{BeniaminiPiran2019}) or addition of other sources of heavy elements (e.g., \citealt{Kobayashietal2020}). \cite{Banerjeeetal2020} argue that taking the natal kicks of binary NSs (with an average velocity of $90 \km \s^{-1}$) and the formation history of the MW into consideration can naturally reproduce the evolution `knee' without the need for neither of them due to the effect of kick induced migration on the frequency and effective delay time distribution of binary NS mergers. We note, however, that other studies find the kick velocities of Galactic binary NSs to be an order of magnitude slower (e.g., \citealt{BeniaminiPiran2016,Taurisetal2017}). Moreover, The abundance pattern predicted by \cite{Banerjeeetal2020} strongly depends on the Galactic radius, an effect which is not consistent with observations. 
Due to all the opposing views (see also  \citealt{Matteuccietal2014, Wehmeyeretal2015, Coteetal2016, KomiyaShigeyama2016, Hotokezakaetal2018, cot19,weh19}), further research is needed to fully resolve the contribution of binary NSs to r-process enrichment in the Galaxy. 

Our study focuses on CEJSNe as a possible r-process site. CEJSNe experience a natal kick from the CCSN that forms the NS in the binary system, driving the binary system away from the explosion site. Since the newly born NS has a massive companion, the natal kick of the binary system is moderate ($\simeq 10-20 \km \s^{-1}$; see \citealt{GrichenerSoker2019b}), leading to some correlation with iron production. In addition, the propagation of the jets dilute the r-process in a large volume. These two effects prevent spatial correlation between the iron production and the later r-process nucleosynthesis. Moreover, the relatively short delay time of CEJSNe allows them to account for r-process production in the early Galaxy and for the evolution `knee' at higher metalicites, as we show next. 

\subsection{The contribution of the CEJSN-r process scenario} 
\label{subsec:results}

We present our results of the Galactic chemical evolution model of the CEJSN r-process scenario (section \ref{sec:Numerics}) by the large-colored dots in Figs. \ref{fig:DelayTimes} and \ref{fig:EuMass}. Our calculations assume that the CEJSNe r-process scenario alone accounts for the formation of europium in the MW. 

In Fig. \ref{fig:DelayTimes} we assume that a single CEJSN r-process event produces europium mass of $M_{\rm Eu}=2\times10^{-5} M_{\rm \odot}$. The three different curves correspond to three different delay times as we indicate in the figure. We find that for ${\rm [Fe/H]} \gtrsim -1$ all curves coincide and reproduce the observed evolution `knee'. Moreover, the possible delay times of the CEJSN r-process scenario (\citealt{GrichenerSoker2019b}) are sufficiently short to produce europium at lower metalicites. The different delay times cover a range of average values of $\rm [Eu/Fe]$ at very low metalicities. The nature of the Galactic chemical evolution models does not allow to study individual events and therefore we do not expect to cover the entire range of stars in the figure. For instance, since in the classical one-zone model of Galactic chemical evolution an instantaneous mixing of the ISM is assumed, we can discuss only the average trend of elemental abundances. In order to reproduce the scatter, it is necessary to take account of small-scale differential mixing. We estimate that in the CEJSN r-process scenario the scatter of elemental abundance ratios would be larger than the case with supernovae only, and smaller than the case with binary NS mergers only.    

In Fig. \ref{fig:EuMass} we present the evolution on the same plane for CEJSNe with a delay time of $t_{\rm d} = 20\Myr$ and three possible values of europium mass per event as we indicate in the figure. All three curves have the same trend. Each curve reproduces the descending evolution `knee' at higher metalicites of ${\rm [Fe/H]} \ga -1$ , and has a more or less constant average value for older stars. We learn that when keeping other parameters unchanged, on average one  CEJSN r-process event cannot produce less than $M_{\rm Eu} \simeq 1.5\times10^{-5} M_{\rm \odot}$ of europium nor more than $M_{\rm Eu} \simeq 2.5\times10^{-5} M_{\rm \odot}$ to reproduce the observations. Note that if we increase the mass of europium per CEJSN r-process event by a factor k, and reduce the number of events by the same factor, we would have obtained the same results since we compute the average value and not the scatter. 

\section{Summary}
\label{sec:Summary}
 
We used a Galactic chemical evolution model (section \ref{sec:Numerics}) to find the contribution of the CEJSN r-process scenario to the europium abundances in the MW. We adopted the parameters of the CEJSN r-process scenario from previous studies \citep{GrichenerSoker2019a, GrichenerSoker2019b} and found that this scenario can account for the observations (Figs. \ref{fig:DelayTimes} and \ref{fig:EuMass}). In particular, we can reproduce the evolution `knee'  together with the delayed iron production from SNe Ia and also the average europium abundance at early times because of the short timescales.

The chemical evolution model studies a large population of events. Therefore, the results in the figures present the average behavior. We note that individual CEJSN r-process events might have a diverse range of properties, which could explain the observed scatter of r-process elements. For instance, the NS-RSG merger that initiates the CEJSN event can take place at different time delays and with different RSG properties, both of which influence the accreted mass onto the NS and therefore the r-process yield per event. 

Some of the parameters that make the events diverse are the amount of iron that the NS progenitor has synthesised, the mass of CO core of the RSG at the time of the event, and the NS mass. 
Moreover, the jets that the NS launches mix with the interstellar medium that surrounds the binary system. The distance the jets propagate depends on the density of the material around the jets and the presence of dense clouds. This leads the r-process material synthesized in the jets to mix with gas that contains different amounts of iron. All these properties influence both the amount of europium and the amount of iron it mixes with, explaining the scatter in the observed ${\rm [Eu/Fe]}$.

Overall, our study demonstrates the advantages of the CEJSN r-process scenario in accounting for the r-process at very early times and in explaining the later evolution. The next step would be to perform hydrodynamical simulations of the NS-core merger, including the launching of jets by the NS and the nucleosynthesis inside the jets including detailed neutrino transport. 

\section*{Acknowledgments}
We thank an anonymous referee for detailed comments that improved our manuscript. This research was supported by a grant from the Israel Science Foundation (769/20).
CK acknowledges funding from the UK Science and Technology Facility Council (STFC) through grant ST/R000905/1 \& ST/V000632/1.
The work was also funded by a Leverhulme Trust Research Project Grant on ``Birth of Elements''.

\end{document}